\pdfoutput=1

\documentclass[letter]{sig-alternate}
\usepackage{graphicx}
\usepackage[english]{babel}
\usepackage{varioref,url}
\usepackage[utf8]{inputenc}
\usepackage{cmap}
\usepackage[T1]{fontenc}
\usepackage{amsmath}
\usepackage{txfonts}
\usepackage{pifont}
\usepackage[protrusion=true, expansion, kerning]{microtype} 
\usepackage{fixltx2e}
\usepackage{tabularx}
\usepackage{newfloat}
\DeclareFloatingEnvironment[fileext=example,placement={htb},name=Example]{example}
\usepackage{xspace}
\newcommand\Dash{---} 

\chardef\bs=`\\

\newcommand{\stress}[1]{\emph{#1}}

\newcommand{\Bpref}{\ensuremath{\textit{Bpref}}\xspace}
\newcommand{\Bprefbf}{\ensuremath{\textit{\textbf{Bpref}}}\xspace}
\renewcommand{\slash}{/\hskip0pt\relax}
\newcommand\Authors{Martin L\'i\v ska, Petr Sojka, Michal R{\r u}{\v z}i{\v c}ka}
\newcommand\Title{Combining Text and Formula Queries in~Math~Information~Retrieval}
\newcommand{\Subtitle}{Evaluation of Query Results Merging Strategies}
\newcommand{\FinalVersionInfo}{%
  ©2015~\Authors.
  This is the author's version of the work. It is posted here for your personal 
  use. Not for redistribution. The definitive version was published in
  proceedings of the First International Workshop on Novel Web Search
  Interfaces and Systems (NWSearch'15), October 23, 2015, Melbourne, Australia:
  The 24th ACM International Conference on Information and Knowledge Management
  (CIKM 2015) Workshop. ACM, New York, NY, USA, 2015. ISBN \mbox{978-1-4503-3789-2.}
  \href{http://dx.doi.org/10.1145/2810355.2810359}{http://dx.doi.org/10.1145/2810355.2810359}.}
\usepackage{hyperref}
\hypersetup{%
  unicode, 
  colorlinks=false, 
  pdftitle={\Title: \Subtitle},
  pdfauthor={\Authors},%
  pdfkeywords={%
      Queries and Query Analysis::query reformulation;
      Queries and Query Analysis::query intent;
      Retrieval Models and Ranking::combining searches;
      Search Engine Architectures and Scalability::efficient querying strategies and tradeoffs;
      IR and Structured Data::XML search;
      Other application::digital libraries;
      Retrieval Models and Ranking::diversity;
      MathML;
      indexing;
      search;
      canonical MathML;
      EuDML;
      information systems;
      information retrieval;
      mathematical content search; 
      math indexing and retrieval (MIR);
      document ranking of mathematical papers},
  pdfsubject={\FinalVersionInfo}}

\toappear{\FinalVersionInfo}

\begin{document}

\clubpenalty=10000
\widowpenalty=10000 

\title{\Title}
\subtitle{\Subtitle}
\numberofauthors{3}
\author{
\alignauthor
Martin L\'i\v ska\\
       \affaddr{Faculty of Informatics, Masaryk University}\\
       \affaddr{Botanick\'a 68a, 602\,00 Brno, Czech Republic}\\
       \email{255768@mail.muni.cz}
\alignauthor
Petr Sojka\\
       \affaddr{Faculty of Informatics, Masaryk University}\\
       \affaddr{Botanick\'a 68a, 602\,00 Brno, Czech Republic}\\
       \email{sojka@fi.muni.cz}
\alignauthor
Michal R{\r u}{\v z}i{\v c}ka\\
       \affaddr{Faculty of Informatics, Masaryk University}\\
       \affaddr{Botanick\'a 68a, 602\,00 Brno, Czech Republic}\\
       \email{mruzicka@mail.muni.cz}
}

\maketitle

\begin{abstract}
Specific to Math Information Retrieval is combining text with
mathematical formulae both in documents and in queries.
Rigorous evaluation of query expansion and merging strategies
combining math and standard textual keyword terms in a query are
given.
It is shown that techniques similar to those known from textual
query processing may be applied in math information retrieval
as well, and lead to a cutting edge performance.
Striping and merging partial results from subqueries is one technique
that improves results measured by information retrieval 
evaluation metrics like \Bpref.
\end{abstract}




\hyphenation{Sojka data-base meta-data}
 
\newcommand\DMLCZ{\mbox{DML-CZ}}
\newcommand{\nth}{\ensuremath{^{\textrm{\scriptsize th}}}}
\newcommand\const{\ensuremath{\mathrm{const}}}
\newcommand\id{\ensuremath{\mathrm{id}}}
\newcommand\formula[1]{\ensuremath{\mathrm{formula}_{#1}}}
\newcommand\term[1]{\ensuremath{\mathrm{term}_{#1}}}
\let\stress\emph
\long\def\TODO#1#2{\par\bgroup{{\bf TODO\\ by #1:}} \it#2\egroup\par}
\def\changemargin#1#2{\list{}{\rightmargin#2\leftmargin#1}\item[]}
\let\endchangemargin=\endlist
\newcommand{\sigirnewpage}{\vskip5\baselineskip\pagebreak[3]\vskip-5\baselineskip\relax}
\def\pmottonosec#1{\par\medskip
 \vbox{{\small\begin{flushright}#1\\ Pablo Picasso\end{flushright}}}}
\def\pmotto#1 #2\section{\par\medskip
 \vbox{{\small\begin{flushright}#2\\ Pablo Picasso\end{flushright}}%
 \vspace*{#1\baselineskip}}\section%
}

\section{Motivation}
There are about 350,000,000 formulae in 
\href{http://www.nature.com/news/the-arxiv-preprint-server-hits-1-million-articles-1.16643}{1,000,000 papers} 
in \href{http://arXiv.org}{arXiv.org}
to be indexed and searched in addition to a keyword-based 
full-text search. Processing of structured objects like
mathematical formulae is not yet supported in production IR systems.
First deployed Math Information Retrieval (MIR) system that 
allowed searching formulae was system~\cite{dml:cicm2014liskaetal:short}
used in the European Digital Mathematical Library~\href{http://eudml.org}{EuDML}.
Math-aware search is now planned on Wikipedia and arXiv.org.
For rigorous evaluation of existing MIR system prototypes 
new Math Tasks have been set up at NTCIR-10 and NTCIR-11 
conferences~\cite{MIR:NTCIR-11}.
There are now datasets and query relevance assessments
available allowing MIR research community to rigorously
evaluate available systems and their ranking strategies. 

In this paper we open a new area of research related
to the query relaxation based on combining math and text
keywords as well as merging results of relaxed subqueries.
Different strategies in detail using datasets
from NTCIR-11 Math Task~2 are evaluated.

Combination of multiple formulae and multiple text keywords 
in one query used in \mbox{NTCIR-11} Math~2 Task~\cite{NTCIR11Math2overview}
seems to be more consistent with the real situation of 
a human using both textual keywords and math formulae to express
search intent. Math formulae are a means how to allow the user to 
filter out relevant documents from the entire database. They are 
complementary to the textual keywords, not the sole way of 
expressing the search intent.

The MIaS system~\cite{dml:doceng2011SojkaLiska}
supports these kind of queries natively. All the keywords are 
posted to the system in one text field\Dash the formulae are written in MathML or 
\TeX{} notation with added dollar signs (\$) on both sides of the \TeX{} 
formulae. Formulae and text keywords are separated by a single space.
The keywords, sometimes consisting of more than one word, are surrounded 
with a single quotation mark (\verb|"|) to handle 
multi-word keywords as a single entity. For experiments 
described in this paper we are using 
open source system MIaS and NTCIR-11 data.

MIaS is a full-text based search system with and extension
for processing mathematical expressions. The formulae from 
documents and queries are canonicalized, expanded to generalized forms to allow
similarity matching, weighted and translated to linear form to be stored
in a full-text index. Documents are ranked with a modified TF-IDF formula
that considers the similarity of the matched formulae.

\section{Complex Query Relaxation}

To increase recall of not very successful queries as well as 
the overall precision, query expansion and resubmission is 
a useful technique. When a user posts a query that finds
no (or very few) results, in order to give at least some 
results to the user albeit with a lower score, the query can be modified
or relaxed and the search run again.
A method to expand a query to multiple queries where each query is 
a subset of the original query consisting of mathematical
and textual terms has been proven to be very helpful~\cite{MIR:NTCIR-11}.
This was the first experiment in this direction in MIR.

Two types of query relaxation are possible. One way is to reduce the number of terms 
if the query consists of more than one term. A combination of reduced terms needs
to be selected, especially if the query consists of text as well as math terms. More query term 
combinations can be run through the system one after another. The important step is then
an effective algorithm for merging result lists with an appropriate weighting. The basic rule
for weighted merging is that the more reduced a query the lower the score
its individual results should get.

Another type of query relaxation is mathematical expression relaxation. If a query
expression is an actual formula with an equal sign, the expression can be split to the left and right
side of the equal sign. These expressions can then form a new query. If the system supports
expressions with wild cards, queries could be relaxed by automatically inserting these. We experimented
with the reduction of the number of terms in the individual text and math parts
of the queries.

\section{LRO Query Expansion}
\label{sec:expansion}

In our approach, the original query consisting of $k$~keywords and $f$ 
formulae is used to generate a set of `subqueries'. At first, the original query 
is used. Then subqueries are generated one by one removing the keywords from the 
query until the query consists of $f$ formulae only. The rest of the subqueries are 
generated with all the keywords and with formulae removed one by 
one until the query consisting of $k$~keywords only is reached. We call this 
expansion method \stress{LRO (Leave Rightmost Out)}.

An example of the complete `subqueries' generation sequence for a query 
consisting of two formulae and three keywords is 
shown in Example~\ref{ex:subqueries-generation}.
 
\begin{example}
  subquery 1 (the original query):\qquad$f_1$\quad$f_2$\quad$k_1$\quad$k_2$\quad$k_3$\\
  subquery 2:\phantom{ (the original query)}\qquad$f_1$\quad$f_2$\quad$k_1$\quad$k_2$\phantom{\quad$k_3$}\\
  subquery 3:\phantom{ (the original query)}\qquad$f_1$\quad$f_2$\quad$k_1$\phantom{\quad$k_2$\quad$k_3$}\\
  subquery 4:\phantom{ (the original query)}\qquad$f_1$\quad$f_2$\phantom{\quad$k_1$\quad$k_2$\quad$k_3$}\\
  subquery 5:\phantom{ (the original query)}\qquad$f_1$\phantom{\quad$f_2$}\quad$k_1$\quad$k_2$\quad$k_3$\\
  subquery 6:\phantom{ (the original query)}\phantom{\qquad$f_1$\quad$f_2$}\quad$k_1$\quad$k_2$\quad$k_3$

\caption{Complete sequence of subqueries derived from the original user's query}
\label{ex:subqueries-generation}
\end{example}

All the subqueries are one by one used to query the system and the partial results lists
are merged (see the next Section) to the final list that 
is presented to the user.

\begin{figure}[tbh]
    \centerline{\hspace*{-7dd}\includegraphics[width=0.5\textwidth]{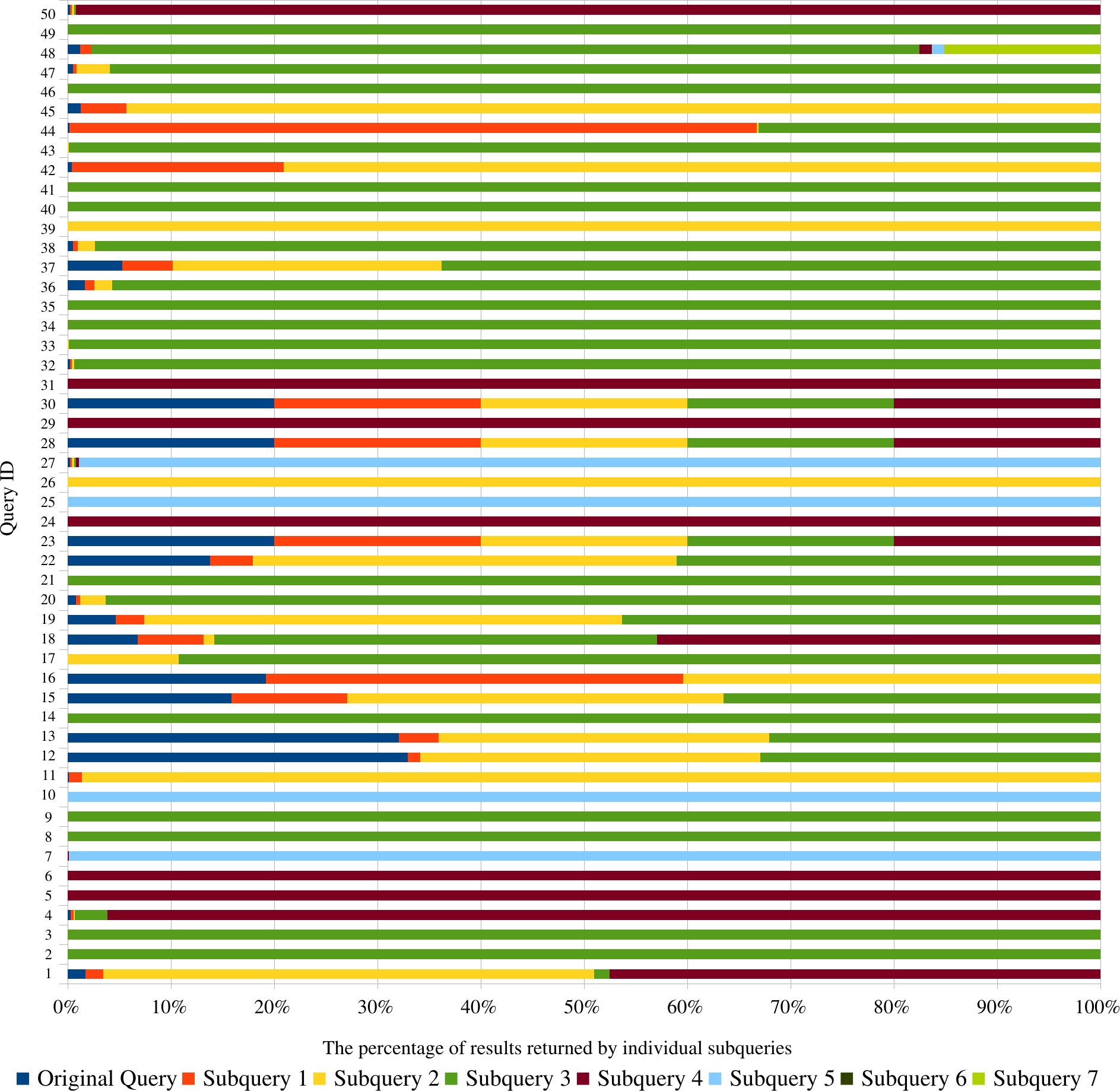}}
    \vspace*{-.8ex}
    \caption{Relative number of results found using different subqueries for
             every query in LRO CMath run}
    \label{fig:cmath-results-per-subquery}
\end{figure}

The statistics of the relative number of results found using each of the subqueries
in CMath run (see Table~\ref{fig:evaluationPCMath}) are shown in
Figure~\ref{fig:cmath-results-per-subquery}. Every subquery was limited to at
most 1,000 results as requested in the \mbox{NTCIR-11} Math Task. The graph
shows that the use of the original unmodified query usually resulted in much
less than requested 1,000 results. The use of the results of multiple
subqueries thus provides significantly more results that are (at least 
partially) relevant to the original topics.

Please note that the last subquery does not contain any formulae,
i.e.\ subquery~6 in Example~\ref{ex:subqueries-generation}, is standard full
text search keyword query with no involvement of mathematical elements
whatsoever.

Please also note that this algorithm does not cover all the possible 
combinations of keywords and formulae as well as `unreasonably' handle different 
formulae differently\Dash in Example~\ref{ex:subqueries-generation} formula~$f_1$ 
is used in five subqueries in contrast to the four uses of~$f_2$
with no reason to prefer $f_1$ before $f_2$. This simplification 
was used to keep the number of subqueries small enough to reach an acceptable 
response time even for interactive real users as the total 
number of subqueries would increase rapidly with the number of formulae and 
keywords in the query if all their possible combinations should be used.
\looseness=-1

The cumulative total MIaS search time for all 50~queries in CMath run was 10.81 
seconds. Cumulative totals for the other three runs are comparable: PMath 12.01\,s, 
PCMath 14.70\,s and for TeX 19.83\,s.

This kind of query expansion provide users with results on more general queries 
than the user originally posted. We consider this behavior useful especially 
for a `research' search as this shows the user a wider context of the query that 
could possibly reveal new and unexpected connections and paths to follow in the 
research. 

\section{Merging of Results}
\label{sec:strip-merging}

The next important step of the query expansion and resubmission procedure is 
the merging of result lists with an appropriate weighting. In conjunction with LRO
expansion method we use a method we refer to as `strip-merging'.

Every subquery results in an ordered list of items with a score%
\footnote{Measure of relevance to the query.}
assigned to each of the results. However, these scores are only comparable 
within the context of their result list. That means that a result $r_1$ with a score of
$0.25$ from the subquery~1 is not necessarily more relevant to the subquery~1 
than a result~$r_2$ with a score of~$0.15$ from the subquery~2 even 
though $0.25 > 0.15$ as absolute scores are incomparable across different subqueries'
results lists. Thus, it is not possible to generate a final results list as 
a simple combination of results from all the subqueries ordered by the score.

Another reason to use a more sophisticated results merging 
procedure is that the results for the 
original query should be preferred to the results found for
subqueries. On the other hand, it is very possible 
that the first result of a subquery could be more relevant
for the user than the 10th result of the original query.

To produce the final results list from the subqueries according to this 
hypothesis we used a method we refer to as `\stress{strip-merging}' of the results. The 
main idea is to interleave the `strips' of hits from all the ordered results 
lists from the subqueries. The less modified subquery to the original query the 
`wider' strip of hits is used in the higher position in the final result list.

Let us have $x$ subqueries (the original one and $x-1$ 
derived subqueries). The top~$x$ most relevant results 
in the final result list are the first~$x$ most 
relevant results from the original query result list, then $x-1$ most
relevant results from the first derived subquery are added, then $x-2$ results 
from the second subquery and so on until the first most relevant result from the
last derived subquery is added. This procedure is then repeated with the next
$x$ results from the results list to the original query, $x-1$ results from the
first subquery etc.\ until the desired amount of results is reached. If all the
results from a subquery are used and  there are no more left we continue without
changing the width of the strips for the other subqueries.

\makeatletter
\def\strategy{\vskip 2pt
  \@startsection{paragraph}{4}{\z@}{3\p@ \@plus \p@}{-5\p@}{\subsecfnt}}
\makeatother

\section{Other Querying Strategies and Result Merging} \label{sec:strategies}

\strategy{Original Query Only: OQO}
\label{sec:oqo}
The basic reference querying strategy is the use of the original query without 
any modifications or derived subqueries. Results found for the original query 
is the final list of results returned to the user.

\strategy{Math Terms Only: MTO}
\label{sec:mto}
\stress{Math Terms Only} querying strategy is simple modification of the 
\stress{Original Query Only} strategy: 
The query consists of formulae from the original query, all the text keywords 
are removed from the query.

\strategy{Text Terms Only: TTO}
\label{sec:tto}
In \stress{Text Terms Only} strategy the query consists of only text
keywords from the original query.

\strategy{All Possible Subqueries: APS}
\label{sec:aps}
The opposite extreme to using only the original query only is to use all 
the possible subqueries derivable from the original query. Provided the original 
query consists of $x$ formulae and $y$ text keywords, all the possible 
combinations of formulae $f_1,
\ldots, f_x$ and text keywords $k_1, \ldots, k_y$ provide us with $2^{x+y}-1$ 
non-empty subqueries (including the original query itself).

Every subquery can be easily identified by a `bit mask' representing the
inclusion\slash exclusion of particular components of the
original query. For example, subquery~5 in 
Example~\ref{ex:subqueries-generation} can be represented with 
mask \stress{10-111}.

The subquery mask can also be used to express importance and degree of 
modification of the particular subquery in contrast to the
original query. We call this number the `\stress{mask weight}'
and it is defined as
$
    \text{mask weight} =
        \sum_x {2 f_x} +
        \sum_y {k_y}
$,
where $f_x$ is value of the $x$-th bit in the formulae part of the mask and 
$k_y$ value of the $y$-th bit in the keywords part.
The value of the formula bit is multiplied by two to 
increase importance of subqueries with maths components.

In the \stress{All Possible Subqueries} querying strategy the final list of 
results is built up from results of particular subqueries as follows:
\begin{enumerate}

    \item Lists of results from all the subqueries are ordered according to their 
        mask weights.

    \item \label{item:work} Let $w_s$ be mask weight of the subquery~$s$.
        For every subquery in the ordered subquery list remove $w_s$ top results 
        from the $s$-th query result list and put them to the final result list.

    \item Repeat Step~\ref{item:work} until all the results were moved to the 
        final result list or a desired number%
        \footnote{We put up to 1,000 results to every final result list.}
        of results in the final result list is reached.
\end{enumerate}

\strategy{Leave One Out: LOO}
\label{sec:loo}
The \stress{Leave One Out} querying strategy is similar to the \stress{All 
Possible Subqueries} strategy with the following differences:
\begin{itemize}
    \item We work with a restricted set of the subqueries\Dash only the original 
        query and derived subqueries with exactly one component (one 
        formula or one text keyword) excluded are used.

    \item In Step~\ref{item:work} of the merging algorithm we do not use mask 
        weight as the `strip-weight'. The strip-weight is
        2 if taking results from the original query results list,
        and 1 otherwise.
\end{itemize} 

Please note that the ordering of the result lists of subqueries with the equal mask 
weight is implementation dependant and not defined.

\strategy{Leave One or Two Out: LOoTO}
\label{sec:looto}
The \stress{Leave One or Two Out} querying strategy is further extension of the 
similar \stress{Leave One Out} strategy:
\begin{itemize}
    \item The set of the subqueries consists of the original query and 
        derived subqueries with exactly one or two components excluded.

    \item The strip-weight is
        3 if taking results from the original query results list,
        2 if taking results from a derived query with exactly one 
            component excluded,
        and 1 otherwise.
\end{itemize}

Once again, the ordering of the result lists of subqueries with the equal mask weight is 
implementation dependant and not defined.

\bgroup
  \catcode`_=\active
  \gdef\works{\catcode`_=\active
      \def_{\setbox0=\hbox{0}\hskip\wd0\relax}}
\egroup

\begin{table*}[tbh]
\vspace*{-.75\baselineskip}
\tabcolsep4dd\centering
\caption{Evaluation metrics for CMath and PMath runs. Values are averaged over 50 NTCIR 
 queries\slash topics. Names of the strategies are described in 
 Sections~\protect\ref{sec:expansion},
 \protect\ref{sec:strip-merging} and \protect\ref{sec:strategies}\Dash
  OQO considered as the baseline. The best value of each metric across the
  strategies is highlighted in bold}
\label{fig:evaluationPCMath}
\smallskip\works
  \begin{tabular}{lc||c|c|c|c|c|c|c}
   metric      & run   &\bf OQO &\bf MTO &\bf TTO &\bf LOO &\bf LOoTO&\bf APS 
   &\bf LRO \\
   \hline\hline
      \Bprefbf & CMath & 0.2544 & 0.2673 & 0.3739 & 0.4623 &  0.4636 & 0.4653 &\bf 0.4734 \\
      \Bprefbf & PMath & 0.2496 & 0.2694 & 0.3739 & 0.448_ &  0.449_ & 0.449_ &\bf 0.4547 \\
   \hline
   \bf MAP avg & CMath & 0.087_ & 0.0879 & 0.1387 & 0.168_ &  0.1479 & 0.1432 &\bf 0.2152 \\
   \bf MAP avg & PMath & 0.0704 & 0.0719 & 0.1387 & 0.1502 &  0.1315 & 0.1252 &\bf 0.1943 \\
   \hline

   \bf P@1 avg & CMath & 0.6667 & 0.6207 & 0.72__ & 0.68__ &  0.62__ & 0.58__ &\bf 0.96__ \\
   \bf P@1 avg & PMath & 0.6538 & 0.6___ & 0.72__ & 0.64__ &  0.58__ & 0.54__ &\bf 0.94__ \\
   \hline

   \bf P@5 avg & CMath & 0.4133 & 0.3793 & 0.604_ & 0.628_ &  0.52__ & 0.516_ &\bf 0.872_ \\
   \bf P@5 avg & PMath & 0.3462 & 0.32__ & 0.604_ & 0.6___ &  0.484_ & 0.456_ &\bf 0.848_ \\
   \hline

   \bf P@10 avg& CMath & 0.27__ & 0.2759 & 0.35__ & 0.432_ &  0.412_ & 0.368_ &\bf 0.546_ \\
   \bf P@10 avg& PMath & 0.2308 & 0.228_ & 0.35__ & 0.406_ &  0.384_ & 0.34__ &\bf 0.506_ \\
  \end{tabular} 
 \end{table*}

\section{Evaluation}
\label{sec:evaluation}

We evaluated the strategies using NTCIR-11 Math-2 Task 
collection of documents and relevance judgements provided
by the conference organizers~\cite{NTCIR11Math2overview}.
The document collection consists of 105,120 scientific documents
from the arXiv pre-print archive. The documents were divided
into 8,301,578 paragraph units. The whole collection contains 59,647,566
mathematical expressions. There are 50 topics (queries)
consisting of one or more math expressions as well as 
one or more textual terms. The judged pool consisted of 
2,501 relevance assessments, ranked from 0 to~4.

In our evaluation we only used binary relevance judgements. 
0~rank for non-relevant, ranks 1--4 for relevant 
e.g.\ \emph{partially relevant} documents according to the 
original NTCIR-11 evaluation.

For the evaluation tool we used a modified version of Terrier's 
evaluation tool~\cite{mir:Ounis:2007:8380}.
The modification resides in added computation of \Bpref metric.
\Bpref is supposed to be more precise than MAP when the judged 
pool is \stress{far from complete}~\cite{mir:craswellBpref},
which is the case for our situation because of the use
NTCIR-11 data relevance assessments.

We evaluated the performance of different query expansion methods
connected with different results merging methods described
in Section~\ref{sec:strategies}. The results are 
summed up in Table~\ref{fig:evaluationPCMath}. 
As baseline we consider OQO column in Table~\ref{fig:evaluationPCMath},
as this is the current state-of-the-art in query expansion in most of the MIR systems.

Two sets of runs were evaluated. They differed in the math notation that
was used for mathematical expressions in queries. Content MathML 
was used for queries in CMath runs and Presentation MathML in PMath runs.
We used these two notations in queries to see, whether they have any 
impact on the usefulness of individual query expansion methods.

In addition to \Bpref, as effectiveness metrics we have used 
Precision at 1, 5, 10 (P@1, P@5, P@10) and Mean Average Precision
(MAP) as they are known in the IR community.

\section{Summary and Conclusions}
Our experiments have shown the importance of query reduction and results
slicing\slash merging techniques in a MIR system like MIaS with 
mixed query sections containing multiple math tokens as 
well as multiple text tokens at the same time.
We use AND logical operator between text keywords
group and math formulae group aiming for
better precision narrowing down the result set of one 
group with the other.

The importance of expansion is underpinned in the evaluation results
of baseline OQO run against all other runs using query expansion
(LRO, LOO, LOoTO, APS). Both \Bpref as well as MAP are
considerably lower than any of the other runs.

The power of the individual parts of the query, e.g.\ math 
and text parts, is shown in the MTO and TTO runs. 
It is interesting to see how separate sections perform 
w.r.t.\ baseline OQO run. This indicates that the original
topic formulation OQO is too restrictive.

From the runs that used query expansion\slash results merging
the LRO run performed the best. It prefers the math part of
the query over the text part. However, from the TTO run we see
that text terms alone retrieve more relevant results than OQO and
the LRO run covers these results as well. This helps when all
the math terms in the query fail, for instance due to the large
complexity of the formula.

As described in Section~\ref{sec:expansion} subqueries are constructed 
by removing the last keyword\slash formula one at a time. This may lead to a 
suspicion, that the success of LRO run resides in the formulation of
the original query\Dash if the terms in the original query were to be
ordered by its significance,
i.e.\ removing the last keyword means removing the least important
keyword which results in a more specific query, differently formulated
queries (i.e.\ with permuted keywords) would fail 
in this strategy. To verify this hypothesis, we created a reversed
original queries. The order of the keywords and formulae were reverted 
in their respective query groups.
The results of these queries with the LRO strategy were roughly the same as
non-reverted original queries. This disproves our hypothesis and 
means that the LRO strategy used with ordered query tokens by
their specificity gives the best results.
Other expansion\slash merging methods yielded slightly worse
evaluation results. 

It is hard to decide whether leaving one or two text or\slash and math tokens 
helps the query performance. It is heavily dependant on the actual terms
in the queries and their restrictiveness.

PMath runs show similar results with slightly lower overall scores.
This is caused by a less precise Presentation MathML query formulae, 
which may contain a semantically less important markup that may lead
to a mismatch between query expression and those found in documents.

\break
Paying attention to query reduction and results slicing is
of utmost importance in MIR. Content MathML gives slightly better
results than Presentation MathML and helps to narrow a semantic gap.

\bibliographystyle{abbrv}

\end{document}